\documentclass[a4paper,11pt]{article}
\usepackage{jheppub}
\newcommand{\cN} {{\cal N}}

\newcommand{\R} {{\mathbb R}}

\newcommand{\cO} {{\cal O}}
\newcommand{\al} {{\alpha}}
\newcommand{\la} {{\lambda}}
\newcommand{\La} {{\Lambda}}

\title{Analytical solutions of pure-spinor superstring field theory}

\author{Michael Kroyter}
\affiliation{School of Physics and Astronomy\\
The Raymond and Beverly Sackler Faculty of Exact Sciences\\
Tel Aviv University, Ramat Aviv, 69978, Israel}
\emailAdd{mikroyt@tau.ac.il}

\abstract{We examine the possibility of constructing analytical solutions
describing marginal deformations in the open superstring field theory that
is based on the non-minimal pure-spinor formalism. It is found out that
some methods used for constructing solutions of bosonic and RNS string
field theories do not seem to generalize to the pure-spinor case,
while other methods do lead to reliable analytical solutions.}

\preprint{MIT-CTP-4183 \begin{flushright} \ \\ \vspace{-8mm} TAUP-2921-10 \end{flushright}}

\begin{document}

\maketitle

\section{Introduction}

Most of the attempts towards a covariant open superstring field
theory~\cite{Witten:1986cc,Witten:1986qs,Preitschopf:1989fc,Arefeva:1989cm,Arefeva:1989cp,Berkovits:1995ab,Berkovits:2000hf,Berkovits:2001im,Arefeva:2002mb,Michishita:2004by,Berkovits:2009gi,Kroyter:2009zj,Kroyter:2012ni,Berkovits:2012np}
are based on the Ramond-Neveu-Schwarz (RNS) world-sheet formulation.
Up to some subtleties, most of these formulations seem to be classically
equivalent when restricted to the NS
sector~\cite{Fuchs:2008zx,Kroyter:2009zi,Kroyter:2009bg,Kroyter:2009zj}.\footnote{See~\cite{Fuchs:2008cc}
for a review of recent developments in string field theory.}
In the Ramond sector, on the other hand, these theories suffer from
various problems~\cite{Wendt:1987zh,Kroyter:2009zi}.
While it seems that these issues can be overcome using
a new (democratic) RNS formulation~\cite{Kroyter:2009rn,Kroyter:2010rk},
it might also be useful
to study superstring field theory in a formulation that treats the NS and
Ramond sectors on the same footing from the beginning, such as the
Green-Schwarz (GS) formalism\footnote{Another possibility would be to
generalize the novel closed superstring field theory of Jur\v co and
M\" unster~\cite{Jurco:2013qra} to the open string case.}.

It is hard to use the GS formalism covariantly due to complications related
to kappa-symmetry. A non-trivial modification of the GS formalism, in which these issues
are resolved, is the pure-spinor formulation of string
theory~\cite{Berkovits:2000fe,Berkovits:2002zk,Berkovits:2001us,Berkovits:2005bt,Berkovits:2007wz}.
A superstring field theory based on the (non-minimal) pure-spinor formulation
exists~\cite{Berkovits:2005bt}.
Since its construction this theory was largely ignored, presumably because
most of the string field theory research following
Sen's conjectures~\cite{Sen:1999mh,Sen:1999xm,Sen:1999nx}
concentrated around tachyon condensation.
Moreover, it was shown that the current pure-spinor formulations
cannot be considered complete and cannot lead to a viable quantum
string field theory, due to problems related to the definition of
the space of string fields~\cite{Aisaka:2008vw,Bedoya:2009np}.
Nonetheless, one would still expect to be able to obtain classical
solutions within this framework, at least solutions that are not
related to a condensation of a tachyon, which is absent.
It is the purpose of this paper to look for analytical solutions of this theory.
Specifically, we study solutions that represent marginal deformations.

The rest of the paper is organized as follows. In section~\ref{sec:PSintro},
we briefly review the non-minimal pure-spinor formalism and the open string
field theory built upon it. Then, in section~\ref{sec:marginal}, we attempt
the construction of solutions describing marginal deformations in
this theory. Some concluding remarks are presented in section~\ref{sec:conc}.

\section{The non-minimal pure-spinor formalism}
\label{sec:PSintro}

The pure-spinor formalism extends the GS formalism, so let us begin by
a brief reminder of the building blocks of the latter.
In this formalism the open string is defined in flat space in terms of the bosonic coordinates
$X^\mu(z)$ $\mbox{($\mu=0..9$)}$ and the Grassmann odd, space-time chiral spinor $\theta^\al(z)$ ($\al=1..16$).
We follow the common practice of the pure-spinor literature and work in the $\al'=2$
convention,
\begin{equation}
\partial X^m(z) \partial X^n(0)\sim -\frac{\eta^{mn}}{z^2}\,.
\end{equation}
The $\partial X^\mu$ are weight one conformal primaries and the $\theta^\al$
are conformal primaries of weight zero.
Central to the theory are the GS constraints.
One can formulate the theory in first-order form with respect to
$\theta^\al$~\cite{Siegel:1985xj}: A spinor $p_\al$, whose chirality
is opposite to that of $\theta^\al$ is added to the system.
This spinor describes the (fermionic) momentum conjugate to
$\theta^\al$. Then, the GS constraints take the form
\begin{equation}
\label{GSconst}
d_\al=p_\al-\frac{1}{2}\partial X^m \gamma_{m\,\al \beta} \theta^\beta
 -\frac{1}{8}(\theta^\gamma \gamma^m_{\gamma\delta} \partial \theta^\delta)
      \gamma_{m\,\al \beta} \theta^\beta\,.
\end{equation}

For the bosonic variables we do not add explicit
momenta variables. The (supersymmetric) bosonic momenta are given by,
\begin{equation}
\label{Pim}
\Pi^m=\partial X^m+\frac{1}{2}\theta^\al \gamma^m_{\al \beta}\partial \theta^\beta\,.
\end{equation}
The $\gamma^m_{\al \beta}$ appearing in~(\ref{GSconst}) and~(\ref{Pim})
(and $\gamma^{m\,\al \beta}$ below) are the symmetric chiral
$16\times 16$ blocks of the gamma matrices in ten dimensions
(``ten dimensional Pauli matrices'').
These matrices obey several identities, which are useful for calculations in
the GS and pure-spinor formalisms. In particular, the following Fierz identities hold
(we are sloppy here with the location of the vector indices, which are contracted in the usual way),
\begin{subequations}
\begin{align}
\label{Fierz}
& \gamma^m_{\al\beta}\gamma^m_{\gamma\delta}+
  \gamma^m_{\al\gamma}\gamma^m_{\beta\delta}+
  \gamma^m_{\al\delta}\gamma^m_{\beta\gamma}=0\,,\\
& \gamma^{pqr}_{\al \beta} \gamma^{qr\,\delta}_{\ \gamma}=
  6\big(\delta_\al^\delta\gamma_{\beta\gamma}^p-
        \delta_\beta^\delta\gamma_{\alpha\gamma}^p\big)+
  2\big(\gamma_{\ \al}^{mp\,\,\delta}\gamma^m_{\beta\gamma}-
        \gamma_{\ \beta}^{mp\,\,\delta}\gamma^m_{\al\gamma}\big)\,,\\
& \gamma^m_{\al\beta}\gamma_{\ \gamma}^{mn\,\,\delta}+
   \gamma^m_{\al\gamma}\gamma_{\ \beta}^{mn\,\,\delta}+
   \gamma^m_{\beta\gamma}\gamma_{\ \al}^{mn\,\,\delta}=
  \gamma^n_{\al\beta}\delta_\gamma^\delta+
  \gamma^n_{\al\gamma}\delta_\beta^\delta+
  \gamma^n_{\beta\gamma}\delta_\al^\delta\,.
\end{align}
\end{subequations}

The minimal pure-spinor formalism was constructed in~\cite{Berkovits:2000fe},
by postulating that the ghost system that should be added to the GS variables
consists	 of an even, zero-weight chiral spinor $\la^\al$
together with its conjugate momentum $w_\al$. The spinor $\la^\al$ is
constrained by the pure-spinor condition,
\begin{equation}
\label{PSLa}
\la^\al \gamma^m_{\al \beta} \la^\beta = 0 \qquad \forall m\,.
\end{equation}
This ad-hoc ghost system together with the GS constraints~(\ref{GSconst})
define the following ad-hoc BRST operator (the subscript $m$
on $Q$ is for ``minimal''),
\begin{equation}
\label{Qm}
Q_m=\frac{1}{2\pi i} \oint dz \la^\al d_\al\,.
\end{equation}
The problems caused by the constraints~(\ref{GSconst}) in the GS formalism do not
occur here due to the pure-spinor condition~(\ref{PSLa}).

There are several reasons for modifying the minimal pure-spinor
formalism, e.g., one cannot devise in this formalism a covariant,
composite $b$-ghost field (except in specific
backgrounds~\cite{Berkovits:2010zz}). Such a field is important
for defining scattering amplitudes and other objects, in light of
the expected identity,
\begin{equation}
Qb(z)=T(z)\,.
\end{equation}
Here, $Q$ is the BRST charge and $T$ is the energy-momentum tensor.
A formalism in which this issue is resolved was
proposed by Berkovits~\cite{Berkovits:2005bt}.
In this formalism a non-minimal sector is added to the conformal fields
of the minimal formalism. The quartet mechanism~\cite{Henneaux:1992ig}
guarantees that this non-minimal sector does not modify the
cohomology of the minimal formalism, which is isomorphic to the
cohomology of the RNS string~\cite{Berkovits:2001us}.
Thus, the non-minimal sector consists of a conjugate pair of even fields
$\bar \la_\al$ and $\bar w^\al$ together with a conjugate pair of odd fields
$r_\al$ and $s^\al$. The field $\bar \la_\al$ can be thought of as the opposite
chirality partner of $\la^\al$. Hence, $\bar \la_\al$ and $r_\al$ must also be
zero-weight primaries. Moreover, they have to obey the
pure-spinor constraints,
\begin{align}
\label{PSbarLa}
\bar\la_\al \gamma^{\al\beta}_m \bar\la_\beta = & 0 \qquad \forall m\,,\\
\label{PSrho}
\bar\la_\al \gamma_m^{\al\beta} r_\beta = & 0 \qquad \forall m\,.
\end{align}
Note, that imposing the condition $r_\al \gamma_m^{\al\beta} r_\beta = 0$
would have no consequences, since it must hold in any case, in light of the odd
character of $r$ and the symmetry property of the $\gamma_m$.
Hence,~(\ref{PSrho}) is used in order to reduce the amount of degrees of
freedom of $r$ to the desired amount.
The decoupling of the quartet ($\bar \la_\al\,, \bar w^\al\,, r_\al\,, s^\al$)
is achieved, as usual, by adding to the minimal BRST charge $Q_m$ the
following non-minimal component,
\begin{equation}
\label{Qnm}
Q_{nm}=\frac{1}{2\pi i} \oint dz \bar w^\al r_\al\,.
\end{equation}
The total BRST charge is defined to be the sum of~(\ref{Qm}) and~(\ref{Qnm}),
\begin{equation}
Q=Q_m+Q_{nm}\,.
\end{equation}

The pure-spinor constraints on $\la^\al$, $\bar \la_\al$ and $r_\al$
lead to gauge symmetries of the respective conjugate momenta.
Observables must depend only on gauge invariant combinations of these
momenta such as
\begin{equation}
\label{gaugeInv4w}
J=w \la\,,\qquad N_{nm}=\frac{1}{2}w \gamma_{nm} \la\,,\qquad T_\la=w\partial \la\,.
\end{equation}
Here, we omitted the contracted spinor indices.
We follow this convention below, e.g.,
we write $\bar \la \la$ instead of $\bar \la_\al \la^\al$.
The gamma matrices with multiple indices, such as the $\gamma_{nm}$ above
are defined as the odd part (i.e., repeated vector indices lead to identically
vanishing expressions) of products of gamma matrices of alternating chirality, i.e.,
\begin{equation}
\begin{aligned}
\label{gammaNM}
(\gamma_{nm})^\al\!_\beta &=\gamma_n^{\al \gamma} \gamma_{m\,\gamma\beta} \qquad & (n\neq m)\,,\\
(\gamma_{nm})_\al\!^\beta &=\gamma_{n\,\al \gamma} \gamma_m^{\gamma\beta} \qquad & (n\neq m)\,.
\end{aligned}
\end{equation}
When indices are omitted we infer the location of the indices from the fields with
which the matrices are contracted, e.g., in the definition of $N_{nm}$ in~(\ref{gaugeInv4w}),
$(\gamma_{nm})^\al\!_\beta$ should be used.

A composite and covariant $b$ field can now be defined,
\begin{align}
\nonumber
b=s \partial \bar \la+& \frac{\bar \la\big(2\Pi^m \gamma_m d
 -N_{mn} \gamma^{mn}\partial \theta-J\partial \theta-\partial^2 \theta
  \big)}{4\bar \la \la}
  +\frac{(\bar \la \gamma^{mnp}r)(d\gamma_{mnp}d+24N_{mn}\Pi_p)}
        {192(\bar \la \la)^2}\\
\label{b}
-&\frac{(r\gamma_{mnp}r)(\bar\la\gamma^m d)N^{np}}{16(\bar\la \la)^3}
  +\frac{(r\gamma_{mnp}r)(\bar\la\gamma^{pqr}r)N^{mn}N_{qr}}
        {128(\bar\la\la)^4}\,.
\end{align}
An interesting characteristic of this expression is its dependence on inverse
powers of $\la$ and $\bar \la$. While this is not a-priori wrong,
due to the fact that $\bar \la \la$ is a zero-weight scalar,
one may wonder whether arbitrary negative powers of $\la$ should be
allowed in defining states (and string fields). In fact the answer
to this question is negative. The simplest way to see this is to note the
existence of the state $\xi(z)$~\cite{Berkovits:2005bt},
\begin{equation}
\xi\equiv \frac{\bar \la \theta}{\bar \la \la+r\theta}
  =\frac{\bar \la \theta}{\bar \la \la}\bigg(1-\frac{r\theta}{\bar \la \la}
	      +\Big(\frac{r\theta}{\bar \la \la}\Big)^2-\ldots\bigg)\,.
\end{equation}
The sum above terminates after the eleventh power of $r$, due to its odd
nature and the pure-spinor constraint~(\ref{PSrho}).
Hence, it has a finite degree of singularity with respect to $\la$.
A direct calculation shows that $\xi$ is a contracting homotopy operator
for $Q$, i.e.,
\begin{equation}
Q\xi=1\,.
\end{equation}
This implies that the cohomology of $Q$ is trivial. Thus, the requirement
of a physical cohomology for $Q$ implies that $\xi$ and similar states
should be kept outside the space of allowed states.
On the other hand, some sort of a singularity with respect to powers of $\la$
should be allowed if $b$, as well as some other ghost
states related to the pure-spinor constraint~\cite{Aisaka:2008vw}, are to be included.
It is not quite clear which condition exactly should one impose on the
space of allowed states.

We are now ready to define the pure-spinor open string field theory action.
This action takes a form,
which is very similar to that of the modified and democratic RNS theories,
\begin{equation}
\label{action}
S=-\int \cN\Big(\frac{1}{2}\Psi Q\Psi+\frac{1}{3}\Psi\Psi\Psi\Big)\,.
\end{equation}
Here, the string fields are multiplied, as usual, using the star product, which we
keep implicit and the regularization factor $\cN$ is a mid-point insertion of the
following zero-weight primary field,
\begin{equation}
\label{cNdef}
\cN=e^{Q\chi}=e^{-\la \bar \la - r \theta}\,,\qquad
\chi\equiv -\bar \la \theta\,.
\end{equation}
While we do not believe that this really makes a difference,
we note that this insertion has a trivial kernel.
The insertion tempers the singularities that originate form zero modes
and enables the derivation from the action of the equations of motion
\begin{equation}
Q\Psi+\Psi\Psi=0\,,
\end{equation}
for string fields with an arbitrary number of $\theta$ insertions.

One could think of candidates for $\chi$ other than the one of~(\ref{cNdef}).
The question then would
be whether the off-shell physics depends on the choice of $\chi$.
In~\cite{Berkovits:2005bt}, it was argued that this is not the case at least
for the simplest possible modification $\chi \rightarrow \rho \chi$, with
$\rho\in \R_+$. We continue for the rest of this paper with the simple
choice for $\chi$~(\ref{cNdef}).

\section{Marginal deformations}
\label{sec:marginal}

In this work we study open string field theory. Hence, the marginal
deformations we are interested in are boundary marginal
deformations~\cite{Recknagel:1998ih}.
Marginal deformations are induced by their vertex operators.
As usual, each vertex operator has two variants, integrated and
unintegrated. There is no $c$-ghost in the pure-spinor formalism and
the two types of vertex operators are related by
\begin{equation}
\label{QUV}
Q U=\partial V,
\end{equation}
where $U$ is the integrated vertex operator (the integrand to be precise)
and $V$ is the unintegrated vertex.
The massless vertex operators take the form,
\begin{align}
\label{V}
V=& \la^\al A_\al\,,\\
\label{genU}
U=& \partial \theta^\al A_\al+\Pi^m B_m +d_\al W^\al+
 \frac{1}{2}N_{mn} F^{mn}\,.
\end{align}
Here $A_\al \equiv A_\al(x,\theta)$ is a superfield obeying the SYM
equations and $B_m$, $W^\al$ and $F_{mn}$ are derived from it in the usual
way.
The case of a constant $A_\al$ can be written as,
\begin{equation}
\label{A}
A_\al=\frac{i}{2}(\gamma^m\theta)_\al a_m
  +\frac{i}{12}(\theta \gamma^{mnp}\theta)(\gamma_{mnp}\chi)_\al=
  \frac{i}{2}(\gamma^m\theta)_\al a_m
  +2i(\chi \gamma^m \theta)(\gamma_m\theta)_\al\,,
\end{equation}
where $a_m$ are the constants parameterizing the photon deformations and
the (Grassmann-odd) $\chi^\al$ are constants that parametrize the photino
deformations ($\chi^\al$ should not be confused with the $\chi$ of~(\ref{cNdef})).
A Fierz identity can be used to relate the two representations
of~(\ref{A}). Plugging~(\ref{A}) into~(\ref{genU}) leads to,
\begin{subequations}
\label{Ugen}
\begin{align}
U\,\,=& \, U_{NS}+U_R\,,\\
\label{U}
U_{NS}= & \, i a_m \partial X^m\,,\\
U_R= & \, 6i \chi^\al q_\al\,,
\end{align}
\end{subequations}
where $q_\al$ are the the supersymmetry currents,
\begin{equation}
\label{qAl}
q_\al=p_\al+\frac{1}{2}\partial X^m (\gamma_m \theta)_\al
 +\frac{1}{24}(\theta \gamma^m \partial \theta)(\gamma_m \theta)_\al\,.
\end{equation}
It is straightforward to see that~(\ref{V}) and~(\ref{Ugen}) indeed
obey~(\ref{QUV}).
In what follows we focus on the NS states~(\ref{U}).
We defer the study of Ramond solutions to future work~\cite{KroyterToAppear}.

The action~(\ref{action}) is very similar to that of the bosonic and
the cubic RNS string field theories. Hence, it seems sensible to look for
solutions of a similar form to those found in these other theories.
In the bosonic theory there are several forms of solutions for marginal
deformations, which we describe below. All these solutions are most
simply represented using insertions over wedge states in the
cylinder coordinates~\cite{Rastelli:2000iu,Schnabl:2005gv}.
The wedge states $W_n$ ($0\leq n\leq \infty$) form a one-parameter family
of zero-ghost-number states. This family interpolates between the identity
string field $W_0$ and the sliver state
$W_\infty$~\cite{Rastelli:2000iu,Rastelli:2001vb,Schnabl:2002gg}.
The state $W_1$ is the perturbative vacuum and the wedge states obey
the simple algebra
\begin{equation}
W_n W_m=W_{n+m}\,.
\end{equation}
We refer the reader to the literature (e.g., section 2 of~\cite{Okawa:2006vm},
or section 4 of~\cite{Fuchs:2008cc})
for more details on these states, the insertions over them and the cylinder coordinates.

The first type of analytical solutions describing marginal deformations
was presented in~\cite{Schnabl:2007az,Kiermaier:2007ba}. There, solutions are
defined in terms of the non-integrated vertex. They include integration over
surface size and it is known how to write them down explicitly only for vertex
operators whose self OPE is regular. These solutions were extended to
cover the NS case in~\cite{Erler:2007rh,Okawa:2007ri,Okawa:2007it}.

A second type of solutions can be used also for vertex operators whose
self-OPE is singular. These solutions do not include integration over
surface size. The solutions can be constructed using one of two approaches.
In the first one, solutions are defined in terms of a formal primitive of the
unintegrated vertex~\cite{Fuchs:2007yy}, while in the second one they are
defined in terms of the integrated vertex~\cite{Kiermaier:2007vu}. The
former approach has the advantage of representing the solution as a (formal)
gauge solution, while the latter is more systematic and might also improve
our general understanding of marginal deformations. The two approaches
can be shown to agree when there are no OPE singularities as well as for some
other solutions~\cite{Kiermaier:2007vu,Fuchs:2008cc}.
Nonetheless, there is no clear proof that this is always the case.
The problem in constructing such a proof is that both approaches
depend on objects that are not defined a-priori. In the first approach
one has to specify the primitive and its OPE with all other fields,
while in the second approach one has to specify the recipe for renormalizing
exponents of integrated vertex operators. Both these concepts rely on
the details of the marginal deformation. Presumably they both hold the same
information and can be properly defined simultaneously. These
solutions were also extended to the NS
case~\cite{Fuchs:2007gw,Kiermaier:2007ki}.

A third type of analytical solutions was presented in~\cite{Kiermaier:2010cf}.
There, the building blocks of the solutions are boundary changing
operators. These solutions are very elegant.
However, as in the case of the first type of solutions,
this construction is defined only for the case of a regular OPE. The extension
of this construction to the NS case was achieved in~\cite{Noumi:2011kn}.

We now wish to examine the possibility to generalize these three methods
to the pure-spinor theory for the test case of the
photon marginal deformation.
First, we attempt, in~\ref{sec:regSol}, to generalize the solutions
related to regular marginal deformations and show that they fail to be
reliable.
Next, in~\ref{sec:SingSol} we examine the generalization of the
solutions that can describe general marginal deformations and show
that a regular generalization does exist.
Then, in~\ref{sec:KMmethod} we examine the case of solutions that
are based on boundary condition changing operators and illustrate
that these solutions also seem not to generalize to the pure-spinor theory.

\subsection{The method for regular marginal deformations}
\label{sec:regSol}

In this method solutions are obtained from the unintegrated vertex.
The photon vertex operator can be read from~(\ref{V}) and~(\ref{A}),
\begin{equation}
\label{unIntVertices}
V=\frac{i a_m}{2}(\la \gamma^m \theta)\,.
\end{equation}
This vertex has regular OPE with itself.
The fact that the OPE of the unintegrated photon
vertex operator with itself is regular, might seem a bit strange, since
this is quite unlike the bosonic case.
OPE singularities might have physical significance, e.g., a
linear divergent term in the self-OPE implies that the deformation
cannot be exactly marginal. However, in the case at hand this obstacle
is absent as the photon deformation is indeed an exactly marginal deformation.

The second building block of the solution is an integral over the
$b$-ghost. As mentioned above, this field is not an ``elementary field''
in the pure-spinor formalism.
Nonetheless, we can attempt to construct the solution using the composite
$b$ field~(\ref{b}).
Specifically, we define, in the cylinder coordinates,
\begin{equation}
B=-\frac{1}{2\pi i}\int_{-i \infty}^{i \infty} b(z) dz\,.
\end{equation}
Being a line integral, $B$ can be deformed, as long
as it does not pass by any other insertion.

The solution of~\cite{Schnabl:2007az,Kiermaier:2007ba}
depends on a marginality parameter $\mu$ and is given by,
\begin{subequations}
\label{regSol}
\begin{align}
\Psi=& \sum_{n=1}^\infty \mu^n\Psi_n\,,\\
\label{initialCond}
 \Psi_1=& \, V(0)W_1\,,\\
\label{PsiNbos}
 \Psi_n=& \int_0^1 dt_1\ldots \int_0^1 dt_{n-1}
     V(0)B V(t_1) B\ldots V(t_1+\ldots t_{n-1})W_{1+\sum t_n}\,.
\end{align}
\end{subequations}
It is easy to note that this defines a solution also in the
pure-spinor case, as long as,
\begin{equation}
\label{VregCond}
V(z) V(0) = \cO(z)\,.
\end{equation}
This is the same condition as in the bosonic case of~\cite{Schnabl:2007az,Kiermaier:2007ba},
since our $V$ is the analogue of $cV$ in these papers. Using~(\ref{unIntVertices}),
one can see that this condition indeed holds.

The solution~(\ref{regSol}) includes an unbounded power of the $b$ ghost
from its appearance in $B$. However, the $b$ ghost~(\ref{b}) includes
negative powers of $\la$ and $\bar \la$. Recall that states including
too high negative powers of $\la$ and $\bar \la$ should be discarded from the space of
string fields. Hence, the $\Psi_n$ above are potentially singular.
In the bosonic case, one can eliminate the
$B$'s, using the properties of the commutator\footnote{Here and elsewhere in this paper
the commutator is a graded one, i.e., for two odd objects, as we have here,
it represents their anticommutator.},
\begin{equation}
\tilde V_{bos}\equiv [B_{bos},V_{bos}]\,.
\end{equation}
The operator $\tilde V_{bos}$ is the matter part of the unintegrated vertex,
\begin{equation}
\label{VcV}
V_{bos}=c \tilde V_{bos}\,,
\end{equation}
and is also equal to the integrated vertex,
\begin{equation}
\label{tildeVisU}
\tilde V_{bos} = U_{bos}\,.
\end{equation}
Then, using the fact that,
\begin{equation}
\label{B2is0}
B_{bos}^2=0\,,	
\end{equation}
which follows from the regularity of the
$b b$ OPE, one can rewrite the solution using no more than a single $B_{bos}$
line integral,
\begin{equation}
\label{wrongSolRep}
\Psi_{n,\,bos}= \int_0^1 dt_1\ldots \int_0^1 dt_{n-1}
     V_{bos}(0) U_{bos}(t_1) U_{bos}(t_1+t_2)\ldots BV_{bos}(t_1+\ldots t_{n-1})W_{1+\sum t_n}\,.
\end{equation}
If we could claim that something similar happens in the
pure-spinor case, the solution~(\ref{regSol}) could be saved.

The $bb$ OPE is regular also for the composite $b$~(\ref{b})~\cite{Chandia:2010ix,Jusinskas:2013}.
This implies that an analogue of~(\ref{B2is0}) holds also in the pure-spinor case.
On the other hand, there is no analogue of~(\ref{VcV}) in this case. In fact,
even for the RNS string, the analogues of~(\ref{VcV}) hold only for specific
pictures~\cite{Friedan:1985ge}. Hence, we have to examine the commutation
relation of $B$ and $V$.
Let us decompose,
\begin{equation}
b=\sum_{k=-1}^3 b_k\,,
\end{equation}
where the $\bar \la$ dependence of $b_k$ is $b_k\sim \bar\la^{-k}$.
Acting on $V$~(\ref{unIntVertices}), the number of $\bar \la$'s does not change.
Hence, the condition that $[B,V]$ has no $\bar \la$ poles is equivalent to the
condition,
\begin{equation}
[B_k,V]=0 \qquad \forall k>0\,.
\end{equation}
Here, we decomposed
\begin{equation}
\label{Bk}
B=\sum_{k=-1}^3 B_k\,,
\end{equation}
and $B_k$ is the component of $B$ for which $b_k$ is the integrand.
However, a direct calculation reveals that,
\begin{equation}
\tilde V_3\equiv [B_3,V]=i\frac{(r\gamma_{mnp}r)(\bar\la\gamma^{pqr}r)
  \big((w\gamma^{mn}\la)(\theta\gamma_s\gamma_{qr}\la)+
       (\theta\gamma_s\gamma^{mn}\la)(w\gamma_{qr}\la)\big)a^s}
        {1024(\bar\la\la)^4}\,.
\end{equation}
We verified that this expression is non-zero for some specific cases
that obey the pure-spinor constraints~(\ref{PSLa}),~(\ref{PSbarLa})
and~(\ref{PSrho}).
This suggests that $\Psi_5$, which has four $B$'s in its definition,
includes a factor of $\bar \la^{-12}$, which cannot be cancelled by
any other factor. This factor is too singular. Hence $\Psi_5$ does not
belong to the space of string fields. This observation renders the
solution itself unphysical.

One could have objected the observation above on several grounds.
First, one might hope that while the $\Psi_n$'s are ``bad'', the resulting
$\Psi$ is still a legitimate string field. This probably cannot be the
case, since $\Psi$ depends on the free parameter $\mu$~(\ref{regSol}).
It is unlikely that the ``magic'' of obtaining a legitimate $\Psi$ from a sum
of the $\Psi_n$'s could occur for all values of $\mu$. Moreover, in the limit
$\mu\rightarrow 0$, the contribution of $\Psi_5$ should become dominant over
that of all the other singular $\Psi_n$'s and as such could not be cancelled.
Alternatively, one could have suggested that we should have considered
not~(\ref{PsiNbos}), but rather~(\ref{wrongSolRep}) for generalizing the
solution in a way that avoids multiple $B$ insertions.
The problem with this would be that~(\ref{wrongSolRep}) does not generalize
to a solution in the pure-spinor case. A direct proof that
this expression defines a solution relies on~(\ref{tildeVisU}), which
does not hold in the pure-spinor case. The closest we can get
in the general case to proving~(\ref{tildeVisU}) is
\begin{equation}
Q(\tilde V-U)=0\,,
\end{equation}
which follows from~(\ref{QUV}) and
\begin{equation}
[K,V]=\partial V\,.
\end{equation}
Here we defined, as usual, the string field $K$ to be an insertion over the identity string
field of the following state, which we also denote by $K$,
\begin{equation}
\label{KQB}
K=QB=-\frac{1}{2\pi i}\int_{-i \infty}^{i \infty} T(z) dz\,.
\end{equation}
Unfortunately, this is not enough, as we explicitly illustrated above.
Yet another reservation might arise from counting the total power of $r$
insertion. For the $\Psi_5$ above it exceeds eleven. Hence, one might think
that while $[B_3,V]$ is non-zero, the part of $\Psi_5$, which is obtained
exclusively from it, is zero, due to the fact that
$r$ is an odd field that obeys the pure-spinor constraint~(\ref{PSrho}).
Such a claim would, however, ignore the fact that the insertions are located at different points.
This means that parts of the expression can have the non-zero modes of $r$
multiplying the inverse power of $\la$ and $\bar \la$ zero modes at a
too-high power. Moreover, we can add one more objection to the solution. The
fact that the vertex operators had regular OPE with themselves was enough
in the bosonic construction, due to the matter-ghost factorization. Here, on
the other hand, it is not clear whether $\tilde V$ has regular OPE with
itself, which might imply that the solution is unreliable even if we do not
care about the issue of $\la$ zero-modes.
In any case, we conclude that solutions of this type do not carry over
in general from the bosonic and NS cases to the pure-spinor case.

\subsection{The method for general marginal deformations}
\label{sec:SingSol}

The second type of solutions can be used even in the case of a marginal
deformation with a singular OPE. While the unintegrated
vertex operator that we consider has regular OPE, the integrated vertex
operator~(\ref{U}), which is the one relevant for this method,
has indeed singular OPE with itself.
As with the previous method~(\ref{regSol}), the solution is given as a power
series in $\mu$, with the same initial condition,
\begin{subequations}
\begin{align}
\label{genSol}
\Psi &= \sum_{n=1}^\infty \mu^n\Psi_n\,,\\
\label{genSolInCond}
 \Psi_1 &= V(0)W_1\,.
\end{align}
\end{subequations}

The integrated vertex operators can be written in terms of their primitives
$\hat U$.
As a result of~(\ref{QUV}), these primitives obey,
\begin{subequations}
\label{prim}
\begin{align}
\partial \hat U &= U\,,\\
Q \hat U &= V\,.
\end{align}
\end{subequations}
In terms of the primitives the solution takes a simple form as a formal gauge
solution~\cite{Fuchs:2007yy},
\begin{equation}
\label{gaugeSol}
\Psi_L=Q\La_L \frac{1}{1-\La_L}\,,
\end{equation}
where $\La_L$ is the formal gauge string field that depends on $\hat U$
and the subscript $L$ stands
for ``left''. There is also a ``right'' solution, which is gauge equivalent
to the left one. A real solution can be obtained by going along this gauge
trajectory to half the way between the two solutions~\cite{Kiermaier:2007vu}.
The main issue with this representation is the
normal ordering of the solution~\cite{Kiermaier:2007vu}
(see also~\cite{Fuchs:2008cc}). However, for the photon deformation this
issue was resolved already in~\cite{Fuchs:2007yy}.
It is obvious that any expression of the form~(\ref{gaugeSol}) is a solution
also in the case of the pure-spinor theory and the main challenge, as in
the other cases, would be to distinguish genuine solutions from pure-gauge
and singular ones.

For the photon, $\hat U$ is the holomorphic half of the $X$ field,
\begin{equation}
\hat U=i a_m X^m(z)\,.
\end{equation}
The gauge string field is,
\begin{subequations}
\label{photonSol}
\begin{align}
\La_L &= \sum_{n=1}^\infty \mu^n \La_n\,,\\
 \La_n &= -\frac{(-i X)^n(0)}{n!}W_n\,,\\
 \label{X}
  X &\equiv a_m X^m\,,
\end{align}
where the insertions are implicitly normal-ordered.
\end{subequations}
Normal ordering $X^n$ can be achieved by point-splitting.
The same results, however, follow from the definition,
\begin{equation}
X^n\equiv \left . \partial_p^n e^{p X} \right |_{p=0}\,,
\end{equation}
with the familiar normal ordering of the exponent.
In the pure-spinor case, this relation implies that,
\begin{equation}
Q X^n= \frac{n (\la \gamma \theta)}{2}X^{n-1}\,,
\end{equation}
where we defined,
\begin{equation}
\gamma\equiv a_m \gamma^m\,.
\end{equation}

From~(\ref{gaugeSol}) and~(\ref{photonSol}) we see that the first few terms
in the expansion~(\ref{genSol}) of the photon solution are,
\begin{subequations}
\begin{align}
\Psi_1 &= Q\La_1=\frac{i}{2} (\la \gamma \theta)(0) W_1\,,\\
\Psi_2 &= Q\La_2+Q\La_1 \La_1=-\frac{(\la \gamma \theta)(0)}{2} 
  \big(X(1)-X(0)\big)W_2\,,\\ \nonumber
\Psi_3 &= Q\La_3+Q\La_2 \La_1+Q\La_1 (\La_2+\La_1 \La_1)\\
       &= -\frac{i(\la \gamma \theta)(0)}{4}
        \big(X(0)^2-2X(0)X(2)+2X(1)X(2)-X(1)X(1)\big)W_3\,.
\end{align}
\end{subequations}
First, we note that $\Psi_1$ indeed agrees with~(\ref{genSolInCond}).
Next, we verify that $\Psi_2$ can be written in terms of the integrated
vertex, i.e., in terms of $\partial X$,
\begin{equation}
\Psi_2=-\frac{(\la \gamma \theta)}{2} \int_0^1 dz \partial X(z) W_2\,.
\end{equation}

With $\Psi_3$ a new complication appears that stems from the fact that
insertions at the same point are normally ordered, but insertions at
different points are not, e.g., by $X(0)^2$ we
mean $:X(0)^2:$, but $X(0)X(2)\neq :X(0)X(2):$. Normal ordering the whole
expressions, we can write the result in terms of $\partial X$ as (implicit
normal ordering),
\begin{equation}
\Psi_3=-\frac{i(\la \gamma \theta)(0)}{4}
        \Bigg(C+\int_0^1\partial X(w)dw
          \Big(\int_0^2+\int_1^2\Big)\partial X(z)dz\Bigg)W_3\,,
\end{equation}
where $C$ is a known constant coming from normal ordering. This constant
depends on the details of the conformal transformation between the
upper half-plane and $W_3$ with the parametrization used for the
location of the insertions. At higher orders various such constants multiply
various powers of $X$, but as explained in~\cite{Fuchs:2007yy}, the result
can always be written in terms of integrals of a normal-ordered polynomial
in $\partial X$.
Employing the method of~\cite{Kiermaier:2007vu} would have lead us, as usual,
exactly to the same results.
We have found out that no pure-spinor-specific complication arise for
this type of solutions. We conclude that this type of solutions adequately
generalizes to the pure-spinor theory.

\subsection{The method based on boundary changing operators}
\label{sec:KMmethod}

In~\cite{Kiermaier:2010cf}, solutions were defined in terms of the boundary
changing operators that change the original BCFT to the one that should be
described by the solution.
This is a desirable description both, because it might be possible to generalized
it to non-marginal deformations and because it relates the space of possible string field
theory solutions to the space of possible BCFTs, which
for the bosonic case should be the same as the space of reliable string backgrounds.
The drawback of this method is that it is, again, appropriate only for
the case of a regular OPE. Since the objects defining the solution are
the boundary changing operators, the regularity condition is defined
in terms of them,
\begin{equation}
\label{regDond}
\sigma_L(z)\sigma_R(0)\sim 1\,.
\end{equation}
Here, $\sigma_L$ is the operator that changes the boundary conditions from
those of the original theory to those of the deformed BCFT and $\sigma_R$
is the operator that changes the boundary conditions back to the original ones.
The fact that the solutions of~\cite{Kiermaier:2010cf} have properties similar to those
that we studied in~\ref{sec:regSol}, is not a coincidence. Indeed, these two types
of solutions are closely related.
However, the functional form of the solution, as introduced in~\cite{Kiermaier:2010cf},
suggests a different type of extension to the pure spinor case, which we now explore.

Let us illustrate how the regularity condition~(\ref{regDond}) is related to the
previously discussed regularity conditions~(\ref{VregCond}).
In the case of a general marginal deformation, the boundary changing operators are defined by
\begin{equation}
\sigma_L(a)\sigma_R(b)=\exp\Big(\mu \int_a^b U(z)dz\Big)\,.
\end{equation}
Here, we kept the dependence of the boundary changing operators on the marginality
parameter $\mu$ implicit.
This expressions should be regularized if $U$ has singularities in its OPE, since
the exponent leads to collisions of the $U$ operators. In the case of a regular OPE
there is no such issue and we can write,
\begin{equation}
\sigma_L(a)\sigma_R(b)=\exp\Big(\mu \big(\hat U(b)-\hat U(a)\big)\Big)=e^{\mu \hat U(b)}e^{-\mu \hat U(a)}\,.
\end{equation}
From here we deduce that the boundary changing operators are the exponents of
the primitives,
\begin{equation}
\label{sigDefHere}
\sigma_L=\sigma_{-\mu}\,,\qquad \sigma_R=\sigma_\mu\,,\qquad \sigma_\mu\equiv e^{\mu \hat U}\,.
\end{equation}
This relation might also hold in the general case of singular OPE, in light of~(\ref{prim}).

Consider now, for simplicity the case of the photon marginal deformation~(\ref{U}).
The integration is immediate and gives,
\begin{equation}
\int_a^b i a_m \partial X^m(z) dz=i a_m \big(X^m(b) - X^m(a)\big)\,.
\end{equation}
As long as $a_m$ is not time-like, one cannot exponentiate this result, due to
OPE singularities. Nonetheless, one can easily guess the properly regularized form
of the boundary changing operators,
\begin{equation}
\label{sigmaOfX}
\sigma_\mu(z)=:e^{i \mu X(z)}:\,.
\end{equation}
Here, we wrote the normal ordering explicitly.
However, while we can define localized boundary changing operators using normal ordering,
the OPE of such objects is non-trivial,
\begin{equation}
\sigma_{\mu_1}(z)\sigma_{\mu_2}(0)=
   z^{a_m a^m \mu_1 \mu_2}\big(\sigma_{\mu_1+\mu_2}(0)+\cO(z)\big)\,,
\end{equation}
and the construction of~\cite{Kiermaier:2010cf} cannot be used.

Note, that the problem we are facing now is quite different from
what we had in~\ref{sec:regSol}. There, the problem seemed to be inherent
and unrelated to the choice of the vector $a_m$.
Here, on the other hand, we can try to continue with the construction, albeit only
for the case of a light-like $a_m$. 
Let us assume then that $a_m$ defines a light-like direction.
In split-string conventions~\cite{Erler:2006hw,Erler:2006ww},
the solution of~\cite{Kiermaier:2010cf} takes the form,
\begin{equation}
\label{KOS1}
\Psi=-\frac{1}{\sqrt{1-K}}Q\sigma_L \frac{1}{1-K} \sigma_R (1-K) B c \frac{1}{\sqrt{1-K}}\,.
\end{equation}
The form~(\ref{KOS1}) is obviously inadequate for generalization to the pure-spinor case,
since it depends on the $c$ ghost, which is absent in this formalism.
Luckily, it was also shown in~\cite{Kiermaier:2010cf}, that the solution
can be rewritten in a $c$-independent way,
\begin{equation}
\label{KOS2}
\Psi=-\frac{1}{\sqrt{1-K}}Q\sigma_L \Big(\sigma_R + \frac{B}{1-K}Q\sigma_R
  \Big)\frac{1}{\sqrt{1-K}}\,.
\end{equation}
In order to obtain~(\ref{KOS2}) from~(\ref{KOS1}), one has to use peculiarities of the
bosonic theory. However, the end result~(\ref{KOS2}) is more general, e.g., it can
constitute a solution also in theories for which~(\ref{VcV}) does not hold.

An important question one should examine at this stage is whether~(\ref{KOS2}) constitutes
a legitimate string field, i.e., whether it is independent of the $X$ zero-mode.
While this is obvious in the bosonic case, in light of the matter-ghost factorization,
the presence of $B$ in the current case can potentially lead to problems.
Thus, we should think of~(\ref{KOS2}) as a formal expression that is defined in
an extended space that includes the zero-mode. Then, we should verify that it is
indeed independent of this zero-mode. This resembles the logic we used
in~\ref{sec:SingSol}, following~\cite{Fuchs:2007yy}.
In order to prove the zero-mode-independence of~(\ref{KOS2}), let us introduce
the operator\footnote{I am grateful to the referee for suggesting this approach.}
\begin{equation}
P\equiv \oint \frac{dz}{2\pi i} \partial X\,.
\end{equation}
This operator is the integral of a weight-one primary and is therefore a derivation
of the star product. A string field is zero-mode-independent if and only if it is
annihilated by this operator.
This is indeed the case for~(\ref{KOS2}), in light of the fact that $[P,Q]=0$.

In~\cite{Noumi:2011kn}, it was directly shown that~(\ref{KOS2}) obeys
the equation of motion, using only the relations~(\ref{regDond}),~(\ref{B2is0}),
(\ref{KQB})~and
\begin{equation}
\label{BcomSigma}
[B,\sigma_{L,R}]=0\,,
\end{equation}
as well as the identities that follow from these relations when they are acted upon
by $Q$.
Another advantage of~(\ref{KOS2}) over~(\ref{KOS1}) is that it manifestly
obeys the reality condition of the string field.
One can simplify~(\ref{KOS2}) slightly using a gauge transformation to the form,
\begin{equation}
\Psi=-\frac{1}{1-K}Q\sigma_L \Big(\sigma_R + \frac{B}{1-K}Q\sigma_R  \Big)\,,
\end{equation}
which does not obey the reality condition. However, the fact that this expression
is gauge equivalent to~(\ref{KOS2}) guarantees that it is also physical.

The expression~(\ref{KOS2}) is linear with respect to $B$. Hence, powers of $\la$ and
$\bar \la$ that are smaller than $-4$ do not appear in it. Thus, it does not suffer
from the problems of the solutions discussed in~\ref{sec:regSol} and could
be considered a legitimate string field.
The only thing one still has to show in order to infer that~(\ref{KOS2}) is
a genuine solution in the case of the pure-spinor theory is
to prove that~(\ref{BcomSigma}) holds in this case, assuming that $\sigma_{L,R}$
is given by~(\ref{sigDefHere}) and $a_m$ a light-like vector.

In the bosonic case the relation~(\ref{BcomSigma}) holds since 
the integrated vertex operator is a pure matter state.
The integrated vertex operator is related to the unintegrated vertex operator via~(\ref{VcV}).
Thus, the latter has ghost dependence. 
However, the boundary changing operator is defined in terms of the integrated vertex
and is, therefore, also a pure-matter state.
In the RNS case~(\ref{VcV}) holds only for specific pictures. Nonetheless, the integrated
vertex operator and the boundary changing operators can be assumed to live in the
matter sector and the relation~(\ref{BcomSigma}) still holds.
Unfortunately, it seems that~(\ref{BcomSigma}) does not hold in general
in the pure-spinor case.
The reason for that is the fact that the $b$ field~(\ref{b}) is a composite field,
which has a non-trivial dependence on the matter fields via its dependence
on~$d_\al$ and~$\Pi^m$.

Let us check then, whether~(\ref{BcomSigma}) holds in the specific case of the photon
deformation. Recall the decomposition
of $B$~(\ref{Bk}). Inspecting the expressions for~$d_\al$~(\ref{GSconst}) and~$\Pi^m$~(\ref{Pim}),
we conclude that each of the $B_k$ should separately commute with the boundary
changing operators. For $B_{-1}$ and~$B_3$ commutation is trivial, since they
do not depend on~$d_\al$ and~$\Pi^m$. The other three components do depend
on~$d_\al$ and~$\Pi^m$. Of these, the dependence of $B_2$ is the simplest
and leads to
\begin{equation}
[B_2,\sigma_\mu]=-\mu \sigma_\mu \frac{(r\gamma_{mnp}r)(\bar\la\gamma^m \gamma \theta)N^{np}}{32(\bar\la \la)^3}\neq 0\,.
\end{equation}
Hence, we conclude that in the case of the photon deformation this type of solutions
cannot be generalized to the pure-spinor case.
Moreover, devising a deformation that commutes with $B$ is probably a non-trivial task.
Thus, the problems that we face in the concrete example we examined are probably quite general.
Note, that similar problems could have occurred also with the first type of solutions
had we been trying to generalize other standard representations of that approach.
In that case we managed to overcome this difficulty only to get to the other problem
that we described. Here, we did not manage to pass even the first obstacle.
Of course, this does not mean that it is impossible to find other generalizations
of these solutions that do work. We are currently studying such
possibilities~\cite{KroyterToAppear}.

\section{Conclusions}
\label{sec:conc}

In this paper we examined the possibility of generalizing known
string field theory solutions to the pure-spinor string field theory,
focusing for simplicity on specific marginal deformations.
While it was found that some solutions seem not to generalize properly,
others had no such problems.
It would be interesting to find more solutions to the pure-spinor string
field theory as well as to further study solutions of this theory by, e.g.,
constructing their boundary states, along the lines
of~\cite{Kiermaier:2008qu,Kudrna:2012re}.

We believe that the existence of solutions to the pure-spinor string field theory
suggests that a reliable pure-spinor string field theory might exist.
Such a theory would generalize the current formalism and would also support
the current solutions.
Furthermore, the results obtained here point towards a specific problem
with the current formulation of pure-spinor string field theory, namely
the definition of the $b$-ghost. Indeed, the solution that generalized
nicely to the pure-spinor case is $b$-ghost independent, while the solutions
that do depend on the $b$-ghost had problems with their generalizations.
Moreover, the problems we faced did not depend on regularity of the OPE
or on any other property that was unrelated to the nature of the composite
$b$-ghost.
We can identify the two problematic characteristics of this field as
its dependence on inverse powers of the pure-spinors $\la$ and $\bar \lambda$
and its dependence on the matter sector, i.e., the lack of matter-ghost
factorization in the pure-spinor formalism.

Of these two problems, the first one stood already behind the claim that
the space of string fields cannot be defined in this framework~\cite{Aisaka:2008vw}.
This is not the ``standard problem'' with defining this space,
which is related to the infinite summation of basis states and the
lack of a natural norm on this space that leads, e.g., to the introduction
of phantom terms~\cite{Schnabl:2005gv,Okawa:2006vm,Fuchs:2006hw,Erler:2007xt}.
Rather, the problem is with defining the basis states themselves.
This problem is present already at the level of closing the algebra of vertex operators.
More precisely, it was shown in~\cite{Aisaka:2008vw} that one has to
include vertex operators with negative powers of $\bar \la \la$, in order
to properly obtain all the needed ghost fields. While this might be acceptable
from the point of view of the first quantized theory, the requirement of
closure under the star product seems to imply that arbitrarily negative
powers of $\bar \la \la$ exist as basis elements of the space of string
fields~\cite{Bedoya:2009np}. The resolution of this problem
presumably calls for a further refinement of the definition of the
pure-spinor worldsheet theory.

We should recall at this point that a-priori we do not really need a $b$-ghost
for the study of marginal deformations. Indeed, we did manage to see that the
most general form of solutions, in which the $b$-ghost is absent,
does generalize to the pure-spinor case.
Moreover, these general solutions do not depend on the non-minimal sector
in an essential way. Their only dependence on the non-minimal sector
comes from their dependence on the energy-momentum tensor.
While we do not have a string field theory action for the minimal pure-spinor theory,
the equations of motion are known and are identical to those of the non-minimal theory
with which we worked so far.
Thus, the solutions that do generalize are automatically solutions for the minimal case as well.
The fact that the solutions do not depend on the non-minimal sector could have been
expected. One could further wonder whether the solutions that did not generalize properly
could be gauge transformed to a regular form that does not depend on the non-minimal
sector. On the one hand, this is problematic, since these solutions are improper from
the point of view of the pure-spinor theory. On the other hand, in the case of the
bosonic theory, the solutions
that did generalize are gauge equivalent to the ones that did not. It is not
that we are missing some solutions, rather we found out that some of the
approaches towards the construction of solutions do not work in the pure-spinor theory.
The solutions that did not work just happened to depend on the $b$-ghost,
which is problematic for other reasons as well.
A regularization of the $b$-ghost was proposed in the context of scattering
amplitudes~\cite{Berkovits:2006vi}. It would be interesting to examine the possibility
of using this regularized $b$-ghost for constructing solutions, as well as for
evaluating amplitudes within string field theory.
However, even if this is possible, it still seems to us that a modified
world-sheet formulation is in order.

Several attempts have been carried out towards the refinement of the
pure-spinor formalism. In particular, Berkovits showed that an elementary
$bc$ ghost system can be introduced into the pure-spinor formalism,
by adding some non-minimal sectors~\cite{Berkovits:2007wz}.
In this formulation the study of tachyon condensation (and more generally
the study of the GSO$(-)$ sector) can be attempted.
Such a formulation could also potentially resolve the problems we faced
with the solutions that did not generalize nicely.
The non-minimal sector of~\cite{Berkovits:2005bt}, considered also
in the current work, is a different one.
It would be interesting to combine the non-minimal sectors
of~\cite{Berkovits:2005bt} and~\cite{Berkovits:2007wz} and to extend the
pure-spinor string field theory to this case.
One would then face the problem of having both a composite and an elementary
$b$ field. This difficulty could be avoided by declaring that negative
powers of $\la$ and $\bar \la$ should not be allowed.
Such a definition could also resolve those problems with defining the space of
string fields that are peculiar to the pure-spinor formalism.
The non-minimal sector of~\cite{Berkovits:2005bt} would still be useful
for obtaining a non-degenerate inner product and for the regularization
and definition of a string field theory action. It might also help
with obtaining a manifestly Lorentz invariant formulation of the theory.
While we feel that such a combined formulation would still be too naive,
it might shed some light on the nature of the sought after proper extension of
the pure-spinor formalism.

\acknowledgments

I would like to thank Ido Adam, Yuri Aisaka, Ever Aldo Arroyo,
Nathan Berkovits, Ted Erler, Stefan Fredenhagen, William Linch, Carlo Maccaferri,
Carlos Mafra, Luca Mazzucato, Yaron Oz, Joris Raeymaekers, Martin Schnabl,
Shingo Torii, Brenno Vallilo, Shimon Yankielowicz and Barton Zwiebach for discussions.
I am especially indebted to Michael Kiermaier for discussion and
for initial collaboration on this project.

Part of this work was performed during my stay at MIT.
While at MIT, this work was supported by the U.S. Department of Energy
(D.O.E.) under cooperative research agreement DE-FG0205ER41360.
This research was supported by an Outgoing International Marie
Curie Fellowship of the European Community. The views presented in this work
are those of the author and do not necessarily reflect those of the European
Community.

\bibliography{bib}

\providecommand{\href}[2]{#2}\begingroup\raggedright\begin{thebibliography}{10}

\bibitem{Witten:1986cc}
E.~Witten, {\it Noncommutative geometry and string field theory},  {\em Nucl.
  Phys.} {\bf B268} (1986) 253.

\bibitem{Witten:1986qs}
E.~Witten, {\it Interacting field theory of open superstrings},  {\em Nucl.
  Phys.} {\bf B276} (1986) 291.

\bibitem{Preitschopf:1989fc}
C.~R. Preitschopf, C.~B. Thorn, and S.~A. Yost, {\it Superstring field theory},
   {\em Nucl. Phys.} {\bf B337} (1990) 363--433.

\bibitem{Arefeva:1989cm}
I.~Y. Arefeva, P.~B. Medvedev, and A.~P. Zubarev, {\it Background formalism for
  superstring field theory},  {\em Phys. Lett.} {\bf B240} (1990) 356--362.

\bibitem{Arefeva:1989cp}
I.~Y. Arefeva, P.~B. Medvedev, and A.~P. Zubarev, {\it New representation for
  string field solves the consistency problem for open superstring field
  theory},  {\em Nucl. Phys.} {\bf B341} (1990) 464--498.

\bibitem{Berkovits:1995ab}
N.~Berkovits, {\it Super-{P}oincar\'e invariant superstring field theory},
  {\em Nucl. Phys.} {\bf B450} (1995) 90--102,
  [\href{http://xxx.lanl.gov/abs/hep-th/9503099}{{\tt hep-th/9503099}}].

\bibitem{Berkovits:2000hf}
N.~Berkovits, A.~Sen, and B.~Zwiebach, {\it Tachyon condensation in superstring
  field theory},  {\em Nucl. Phys.} {\bf B587} (2000) 147--178,
  [\href{http://xxx.lanl.gov/abs/hep-th/0002211}{{\tt hep-th/0002211}}].

\bibitem{Berkovits:2001im}
N.~Berkovits, {\it The {R}amond sector of open superstring field theory},  {\em
  JHEP} {\bf 11} (2001) 047,
  [\href{http://xxx.lanl.gov/abs/hep-th/0109100}{{\tt hep-th/0109100}}].

\bibitem{Arefeva:2002mb}
I.~Y. Arefeva, D.~M. Belov, and A.~A. Giryavets, {\it Construction of the
  vacuum string field theory on a non-{BPS} brane},  {\em JHEP} {\bf 09} (2002)
  050, [\href{http://xxx.lanl.gov/abs/hep-th/0201197}{{\tt hep-th/0201197}}].

\bibitem{Michishita:2004by}
Y.~Michishita, {\it A covariant action with a constraint and {F}eynman rules
  for fermions in open superstring field theory},  {\em JHEP} {\bf 01} (2005)
  012, [\href{http://xxx.lanl.gov/abs/hep-th/0412215}{{\tt hep-th/0412215}}].

\bibitem{Berkovits:2009gi}
N.~Berkovits and W.~Siegel, {\it Regularizing cubic open {Neveu-Schwarz} string
  field theory},  {\em JHEP} {\bf 11} (2009) 021,
  [\href{http://xxx.lanl.gov/abs/0901.3386}{{\tt arXiv:0901.3386}}].

\bibitem{Kroyter:2009zj}
M.~Kroyter, {\it On string fields and superstring field theories},  {\em JHEP}
  {\bf 0908} (2009) 044, [\href{http://xxx.lanl.gov/abs/0905.1170}{{\tt
  arXiv:0905.1170}}].

\bibitem{Kroyter:2012ni}
M.~Kroyter, Y.~Okawa, M.~Schnabl, S.~Torii, and B.~Zwiebach, {\it {Open
  superstring field theory I: gauge fixing, ghost structure, and propagator}},
  {\em JHEP} {\bf 1203} (2012) 030,
  [\href{http://xxx.lanl.gov/abs/1201.1761}{{\tt arXiv:1201.1761}}].

\bibitem{Berkovits:2012np}
N.~Berkovits, {\it Constrained {BV} description of string field theory},  {\em
  JHEP} {\bf 1203} (2012) 012, [\href{http://xxx.lanl.gov/abs/1201.1769}{{\tt
  arXiv:1201.1769}}].

\bibitem{Fuchs:2008zx}
E.~Fuchs and M.~Kroyter, {\it On the classical equivalence of superstring field
  theories},  {\em JHEP} {\bf 0810} (2008) 054,
  [\href{http://xxx.lanl.gov/abs/0805.4386}{{\tt arXiv:0805.4386}}].

\bibitem{Kroyter:2009zi}
M.~Kroyter, {\it Superstring field theory equivalence: {Ramond sector}},  {\em
  JHEP} {\bf 0910} (2009) 044, [\href{http://xxx.lanl.gov/abs/0905.1168}{{\tt
  arXiv:0905.1168}}].

\bibitem{Kroyter:2009bg}
M.~Kroyter, {\it Comments on superstring field theory and its vacuum solution},
   {\em JHEP} {\bf 0908} (2009) 048,
  [\href{http://xxx.lanl.gov/abs/0905.3501}{{\tt arXiv:0905.3501}}].

\bibitem{Fuchs:2008cc}
E.~Fuchs and M.~Kroyter, {\it Analytical solutions of open string field
  theory},  {\em Phys.Rept.} {\bf 502} (2011) 89--149,
  [\href{http://xxx.lanl.gov/abs/0807.4722}{{\tt arXiv:0807.4722}}].

\bibitem{Wendt:1987zh}
C.~Wendt, {\it Scattering amplitudes and contact interactions in {W}itten's
  superstring field theory},  {\em Nucl. Phys.} {\bf B314} (1989) 209.

\bibitem{Kroyter:2009rn}
M.~Kroyter, {\it Superstring field theory in the democratic picture},  {\em
  Adv.Theor.Math.Phys.} {\bf 15} (2011) 741,
  [\href{http://xxx.lanl.gov/abs/0911.2962}{{\tt arXiv:0911.2962}}].

\bibitem{Kroyter:2010rk}
M.~Kroyter, {\it Democratic superstring field theory: {Gauge} fixing},  {\em
  JHEP} {\bf 1103} (2011) 081, [\href{http://xxx.lanl.gov/abs/1010.1662}{{\tt
  arXiv:1010.1662}}].

\bibitem{Jurco:2013qra}
B.~Jur\v{c}o and K.~{M\"{u}nster}, {\it Type {II} superstring field theory:
  Geometric approach and operadic description},
  \href{http://xxx.lanl.gov/abs/1303.2323}{{\tt arXiv:1303.2323}}.

\bibitem{Berkovits:2000fe}
N.~Berkovits, {\it {Super-Poincare} covariant quantization of the superstring},
   {\em JHEP} {\bf 04} (2000) 018,
  [\href{http://xxx.lanl.gov/abs/hep-th/0001035}{{\tt hep-th/0001035}}].

\bibitem{Berkovits:2002zk}
N.~Berkovits, {\it {ICTP} lectures on covariant quantization of the
  superstring},  \href{http://xxx.lanl.gov/abs/hep-th/0209059}{{\tt
  hep-th/0209059}}.

\bibitem{Berkovits:2001us}
N.~Berkovits, {\it Relating the {RNS} and pure spinor formalisms for the
  superstring},  {\em JHEP} {\bf 08} (2001) 026,
  [\href{http://xxx.lanl.gov/abs/hep-th/0104247}{{\tt hep-th/0104247}}].

\bibitem{Berkovits:2005bt}
N.~Berkovits, {\it Pure spinor formalism as an {N = 2} topological string},
  {\em JHEP} {\bf 10} (2005) 089,
  [\href{http://xxx.lanl.gov/abs/hep-th/0509120}{{\tt hep-th/0509120}}].

\bibitem{Berkovits:2007wz}
N.~Berkovits, {\it Explaining the pure spinor formalism for the superstring},
  {\em JHEP} {\bf 01} (2008) 065,
  [\href{http://xxx.lanl.gov/abs/0712.0324}{{\tt arXiv:0712.0324}}].

\bibitem{Sen:1999mh}
A.~Sen, {\it Descent relations among bosonic {D}-branes},  {\em Int. J. Mod.
  Phys.} {\bf A14} (1999) 4061--4078,
  [\href{http://xxx.lanl.gov/abs/hep-th/9902105}{{\tt hep-th/9902105}}].

\bibitem{Sen:1999xm}
A.~Sen, {\it Universality of the tachyon potential},  {\em JHEP} {\bf 12}
  (1999) 027, [\href{http://xxx.lanl.gov/abs/hep-th/9911116}{{\tt
  hep-th/9911116}}].

\bibitem{Sen:1999nx}
A.~Sen and B.~Zwiebach, {\it Tachyon condensation in string field theory},
  {\em JHEP} {\bf 03} (2000) 002,
  [\href{http://xxx.lanl.gov/abs/hep-th/9912249}{{\tt hep-th/9912249}}].

\bibitem{Aisaka:2008vw}
Y.~Aisaka, E.~A. Arroyo, N.~Berkovits, and N.~Nekrasov, {\it Pure spinor
  partition function and the massive superstring spectrum},  {\em JHEP} {\bf
  08} (2008) 050, [\href{http://xxx.lanl.gov/abs/0806.0584}{{\tt
  arXiv:0806.0584}}].

\bibitem{Bedoya:2009np}
O.~A. Bedoya and N.~Berkovits, {\it {GGI} lectures on the pure spinor formalism
  of the superstring},  \href{http://xxx.lanl.gov/abs/0910.2254}{{\tt
  arXiv:0910.2254}}.

\bibitem{Siegel:1985xj}
W.~Siegel, {\it Classical superstring mechanics},  {\em Nucl.Phys.} {\bf B263}
  (1986) 93.

\bibitem{Berkovits:2010zz}
N.~Berkovits and L.~Mazzucato, {\it Taming the $b$ antighost with
  {Ramond-Ramond flux}},  {\em JHEP} {\bf 1011} (2010) 019,
  [\href{http://xxx.lanl.gov/abs/1004.5140}{{\tt arXiv:1004.5140}}].

\bibitem{Henneaux:1992ig}
M.~Henneaux and C.~Teitelboim, {\it Quantization of gauge systems}, .
  Princeton, USA: Univ. Pr. (1992) 520 p.

\bibitem{Recknagel:1998ih}
A.~Recknagel and V.~Schomerus, {\it Boundary deformation theory and moduli
  spaces of {D}-branes},  {\em Nucl. Phys.} {\bf B545} (1999) 233--282,
  [\href{http://xxx.lanl.gov/abs/hep-th/9811237}{{\tt hep-th/9811237}}].

\bibitem{KroyterToAppear}
M.~Kroyter, {\it Work in progress}, .

\bibitem{Rastelli:2000iu}
L.~Rastelli and B.~Zwiebach, {\it Tachyon potentials, star products and
  universality},  {\em JHEP} {\bf 09} (2001) 038,
  [\href{http://xxx.lanl.gov/abs/hep-th/0006240}{{\tt hep-th/0006240}}].

\bibitem{Schnabl:2005gv}
M.~Schnabl, {\it Analytic solution for tachyon condensation in open string
  field theory},  {\em Adv. Theor. Math. Phys.} {\bf 10} (2006) 433--501,
  [\href{http://xxx.lanl.gov/abs/hep-th/0511286}{{\tt hep-th/0511286}}].

\bibitem{Rastelli:2001vb}
L.~Rastelli, A.~Sen, and B.~Zwiebach, {\it Boundary {CFT} construction of
  {D}-branes in vacuum string field theory},  {\em JHEP} {\bf 11} (2001) 045,
  [\href{http://xxx.lanl.gov/abs/hep-th/0105168}{{\tt hep-th/0105168}}].

\bibitem{Schnabl:2002gg}
M.~Schnabl, {\it Wedge states in string field theory},  {\em JHEP} {\bf 01}
  (2003) 004, [\href{http://xxx.lanl.gov/abs/hep-th/0201095}{{\tt
  hep-th/0201095}}].

\bibitem{Okawa:2006vm}
Y.~Okawa, {\it Comments on {S}chnabl's analytic solution for tachyon
  condensation in {W}itten's open string field theory},  {\em JHEP} {\bf 04}
  (2006) 055, [\href{http://xxx.lanl.gov/abs/hep-th/0603159}{{\tt
  hep-th/0603159}}].

\bibitem{Schnabl:2007az}
M.~Schnabl, {\it Comments on marginal deformations in open string field
  theory},  {\em Phys. Lett.} {\bf B654} (2007) 194--199,
  [\href{http://xxx.lanl.gov/abs/hep-th/0701248}{{\tt hep-th/0701248}}].

\bibitem{Kiermaier:2007ba}
M.~Kiermaier, Y.~Okawa, L.~Rastelli, and B.~Zwiebach, {\it Analytic solutions
  for marginal deformations in open string field theory},  {\em JHEP} {\bf 01}
  (2008) 028, [\href{http://xxx.lanl.gov/abs/hep-th/0701249}{{\tt
  hep-th/0701249}}].

\bibitem{Erler:2007rh}
T.~Erler, {\it Marginal solutions for the superstring},  {\em JHEP} {\bf 07}
  (2007) 050, [\href{http://xxx.lanl.gov/abs/0704.0930}{{\tt 0704.0930}}].

\bibitem{Okawa:2007ri}
Y.~Okawa, {\it Analytic solutions for marginal deformations in open superstring
  field theory},  {\em JHEP} {\bf 09} (2007) 084,
  [\href{http://xxx.lanl.gov/abs/0704.0936}{{\tt 0704.0936}}].

\bibitem{Okawa:2007it}
Y.~Okawa, {\it Real analytic solutions for marginal deformations in open
  superstring field theory},  {\em JHEP} {\bf 09} (2007) 082,
  [\href{http://xxx.lanl.gov/abs/0704.3612}{{\tt 0704.3612}}].

\bibitem{Fuchs:2007yy}
E.~Fuchs, M.~Kroyter, and R.~Potting, {\it Marginal deformations in string
  field theory},  {\em JHEP} {\bf 0709} (2007) 101,
  [\href{http://xxx.lanl.gov/abs/0704.2222}{{\tt arXiv:0704.2222}}].

\bibitem{Kiermaier:2007vu}
M.~Kiermaier and Y.~Okawa, {\it Exact marginality in open string field theory:
  a general framework},  {\em JHEP} {\bf 11} (2009) 041,
  [\href{http://xxx.lanl.gov/abs/0707.4472}{{\tt arXiv:0707.4472}}].

\bibitem{Fuchs:2007gw}
E.~Fuchs and M.~Kroyter, {\it Marginal deformation for the photon in
  superstring field theory},  {\em JHEP} {\bf 0711} (2007) 005,
  [\href{http://xxx.lanl.gov/abs/0706.0717}{{\tt arXiv:0706.0717}}].

\bibitem{Kiermaier:2007ki}
M.~Kiermaier and Y.~Okawa, {\it General marginal deformations in open
  superstring field theory},  {\em JHEP} {\bf 11} (2009) 042,
  [\href{http://xxx.lanl.gov/abs/0708.3394}{{\tt arXiv:0708.3394}}].

\bibitem{Kiermaier:2010cf}
M.~Kiermaier, Y.~Okawa, and P.~Soler, {\it Solutions from boundary condition
  changing operators in open string field theory},  {\em JHEP} {\bf 1103}
  (2011) 122, [\href{http://xxx.lanl.gov/abs/1009.6185}{{\tt
  arXiv:1009.6185}}].

\bibitem{Noumi:2011kn}
T.~Noumi and Y.~Okawa, {\it Solutions from boundary condition changing
  operators in open superstring field theory},  {\em JHEP} {\bf 1112} (2011)
  034, [\href{http://xxx.lanl.gov/abs/1108.5317}{{\tt arXiv:1108.5317}}].

\bibitem{Chandia:2010ix}
O.~Chandia, {\it The $b$ ghost of the pure spinor formalism is nilpotent},
  {\em Phys.Lett.} {\bf B695} (2011) 312--316,
  [\href{http://xxx.lanl.gov/abs/1008.1778}{{\tt arXiv:1008.1778}}].

\bibitem{Jusinskas:2013}
R.~L. Jusinskas, {\it Nilpotency of the $b$ ghost in the non minimal pure
  spinor formalism},  \href{http://xxx.lanl.gov/abs/1303.3966}{{\tt
  arXiv:1303.3966}}.

\bibitem{Friedan:1985ge}
D.~Friedan, E.~J. Martinec, and S.~H. Shenker, {\it Conformal invariance,
  supersymmetry and string theory},  {\em Nucl. Phys.} {\bf B271} (1986) 93.

\bibitem{Erler:2006hw}
T.~Erler, {\it Split string formalism and the closed string vacuum},  {\em
  JHEP} {\bf 05} (2007) 083,
  [\href{http://xxx.lanl.gov/abs/hep-th/0611200}{{\tt hep-th/0611200}}].

\bibitem{Erler:2006ww}
T.~Erler, {\it Split string formalism and the closed string vacuum. {II}},
  {\em JHEP} {\bf 05} (2007) 084,
  [\href{http://xxx.lanl.gov/abs/hep-th/0612050}{{\tt hep-th/0612050}}].

\bibitem{Kiermaier:2008qu}
M.~Kiermaier, Y.~Okawa, and B.~Zwiebach, {\it The boundary state from open
  string fields},  \href{http://xxx.lanl.gov/abs/0810.1737}{{\tt
  arXiv:0810.1737}}.

\bibitem{Kudrna:2012re}
M.~Kudrna, C.~Maccaferri, and M.~Schnabl, {\it Boundary state from {Ellwood}
  invariants},  \href{http://xxx.lanl.gov/abs/1207.4785}{{\tt
  arXiv:1207.4785}}.

\bibitem{Fuchs:2006hw}
E.~Fuchs and M.~Kroyter, {\it On the validity of the solution of string field
  theory},  {\em JHEP} {\bf 0605} (2006) 006,
  [\href{http://xxx.lanl.gov/abs/hep-th/0603195}{{\tt hep-th/0603195}}].

\bibitem{Erler:2007xt}
T.~Erler, {\it Tachyon vacuum in cubic superstring field theory},  {\em JHEP}
  {\bf 01} (2008) 013, [\href{http://xxx.lanl.gov/abs/0707.4591}{{\tt
  arXiv:0707.4591}}].

\bibitem{Berkovits:2006vi}
N.~Berkovits and N.~Nekrasov, {\it Multiloop superstring amplitudes from
  non-minimal pure spinor formalism},  {\em JHEP} {\bf 0612} (2006) 029,
  [\href{http://xxx.lanl.gov/abs/hep-th/0609012}{{\tt hep-th/0609012}}].

\end{thebibliography}\endgroup

\end{document}